\documentclass[12pt,reqno]{amsart}
\usepackage{amsmath,amssymb,amsfonts,amsthm}
\usepackage[mathscr]{eucal}
\usepackage[all]{xy}
\usepackage{hyperref}
\usepackage{setspace}
\usepackage{ytableau}
\allowdisplaybreaks

\author{V.A.~Abakumova, S.L.~Lyakhovich}

\address{Physics Faculty, Tomsk State University, Lenin ave. 36, Tomsk 634050, Russia}

\email{abakumova@phys.tsu.ru, \, sll@phys.tsu.ru}

\title{Unfree gauge symmetry}
\begin{document}
\maketitle

\begin{abstract}
The gauge symmetry is said unfree if the gauge transformation leaves the action functional unchanged provided for the gauge parameters are constrained by the system of partial differential equations. The best known example of this phenomenon is the volume preserving diffeomorphism being the gauge symmetry of unimodular gravity (UG). Various extensions are known of the UG, including the higher spin analogs -- all with unfree gauge symmetry. Given the distinctions of the unfree gauge symmetry from the symmetry with unrestricted gauge parameters, the algebra of gauge transformations is essentially different. These distinctions have consequences for all the key constituents of general gauge theory, starting from the second Noether theorem, Hamiltonian constrained formalism, BRST complex, and quantization. In this review article, we summarise the modifications of general gauge theory worked out in recent years to cover the case of unfree gauge symmetry.
\end{abstract}

\label{sec:intro}
\section*{Introduction}

The common textbook definition  of gauge symmetry \cite{Henneaux:1992ig} implies that action functional
is invariant under the gauge variations of the fields
\vspace{-2mm}
\begin{equation}\label{GS}
\delta_\epsilon\phi^i=R^i_\alpha(\phi)\epsilon^\alpha\, , \quad \delta_\epsilon S(\phi)\equiv\epsilon^\alpha R^i_\alpha(\phi) \partial_iS(\phi)\equiv 0 \, ,\quad \forall\,\epsilon^\alpha\,,
\vspace{-2mm}
\end{equation}
where DeWitt's condensed notion is used. The gauge generators $R^i_\alpha(\phi)$ are assumed to be local differential operators which do not vanish on-shell. The gauge parameters $\epsilon^\alpha$ are supposed to be arbitrary functions of space-time. This definition is a cornerstone of general gauge theory, though the examples have been long known of the gauge symmetry that do not fit setup (\ref{GS}).
The deviation from (\ref{GS}) is that gauge variation is unfree in the sense that gauge parameters have to be restricted by the system of partial differential equations (PDE) to leave the action unchanged. One more common assumption of the general gauge theory \cite{Henneaux:1992ig} is that any on-shell vanishing local quantity should reduce to the l.h.s. of field equations and their derivatives. This assumption is also invalid in various known examples of unfree gauge symmetry \cite{Alvarez:2006uu},
\cite{Blas:2007pp},
\cite{Skvortsov:2007kz}, \cite{Campoleoni:2012th}.  The on-shell trivial quantities exist such that do not reduce to the field equations. This general feature of unfree gauge symmetry has been first noticed in \cite{Kaparulin:2019quz}.

Let us first exemplify these general features of unfree gauge symmetry by the case of unimodular gravity (UG) \cite{Buchmuller:1988wx}--\cite{Alvarez:2012px}. For basic introduction into the UG, and further literature, we refer to \cite{Alvarez:2023utn}.
Once the metrics are restricted in UG by the unimodularity condition $\det{g}=-\,1$, gauge symmetry reduces to the volume preserving diffeomorphisms:
\vspace{-2mm}
\begin{equation}\label{Tdiff}
\delta_\epsilon \det{g} = 0\,,\quad \delta_\epsilon
g_{\mu\nu}=\nabla_\mu\epsilon_\nu + \nabla_\nu\epsilon_\mu \quad
\Rightarrow\quad \nabla_\mu \epsilon^\mu=0\,.
\vspace{-2mm}
\end{equation}
Einstein's equations become traceless, hence they are not transverse. This makes $\Lambda$ ``integration constant'', not pre-defined parameter:
\vspace{-2mm}
\begin{equation}\label{Gtilde}
S=\int d^dx\,R\,, \quad \frac{\delta S}{\delta g^{\mu\nu}}\equiv
R_{\mu\nu}-\frac{1}{d}g_{\mu\nu}R\approx 0\,;
\vspace{-2mm}
\end{equation}
\vspace{-2mm}
\begin{equation}\label{UGGI}
 \nabla^\nu
\frac{\delta S}{\delta g^{\mu\nu}}\equiv \frac{d-2}{d}\partial_\mu
R\approx 0 \quad\Rightarrow\quad R\approx \Lambda=const\,.
\vspace{-1mm}
\end{equation}
On-shell relation $R-\Lambda \approx 0$ is \emph{not a differential consequence} of equations of motion (EoM's) (\ref{Gtilde}), nor $\Lambda$ is it a charge of any local conserved current.

Volume preserving diffeomorphisms form the subalgebra
\vspace{-2mm}
\begin{equation}\label{Lie-Tdiff}  \delta_{\epsilon_1}\delta_{\epsilon_2}-\delta_{\epsilon_2}\delta_{\epsilon_1}=
\delta_{[\epsilon_1,\epsilon_2]}\,, \quad
\nabla\cdot\epsilon_{1,2}=0\quad\Rightarrow
\quad\nabla\cdot[\epsilon_1,\epsilon_2]=0\,.
\vspace{-2mm}
\end{equation}
The subalgebra is singled out by imposing PDE  onto the gauge
parameters $\epsilon^\mu$ rather than by explicitly separating
subset of generators.

Various generalizations are known of the UG, see \cite{Barvinsky:2017pmm}-- \cite{Jirousek:2020vhy}. The most frequent starting point for modifications is that unimodularity
condition is replaced by a more general relation, $N=N(\overset{*}{g})$. In this case, $\Lambda$  is still an integration constant (in \cite{Jirousek:2018ago}, \cite{Jirousek:2020vhy} it is Newtonian constant), but there are also new options to describe ``$k$-essence'' and other phenomena.

Higher spin (HS) linearised gravities provide more examples of unfree gauge symmetry. They include irreducible HS traceless tensors \cite{Skvortsov:2007kz},
\vspace{-2mm}
\begin{equation}\label{Sk-V}
    \text{Tr}\, \widetilde{h}=0\, , \quad \delta_{\tilde{\epsilon}}\,
    \widetilde{h}{}_{\mu_1\dots\mu_s}
    =\partial_{(\mu_1}{\widetilde{\epsilon}}_{\mu_2\dots\mu_s)}\,, \quad \text{Tr}\,\widetilde{\epsilon}=0\,,\quad \partial\cdot{\widetilde{\epsilon}} = 0\,,
    \vspace{-2mm}
\end{equation}
as well as ``Maxwell-like'' HS, traceful tensors \cite{Campoleoni:2012th},
\vspace{-2mm}
\begin{equation}\label{C-F}
    \delta_\epsilon h_{\mu_1\dots\mu_s}=\partial_{(\mu_1}\epsilon_{\mu_2\dots\mu_s)}\,,\quad\partial\cdot\epsilon=0\,.
    \vspace{-2mm}
\end{equation}
Both models don't involve auxiliary fields, unlike Fronsdal's action.

HS gravity models with unfree gauge symmetry admit ``global
conserved quantities'', being HS analogs of cosmological constant \cite{Abakumova:2020ajc}, \cite{Abakumova:uzmu}, \cite{Abakumova:2022esw}. Number of these ``HS cosmological constants'' is growing with spin.

For $s=3$, the analog of scalar curvature is a vector.
\vspace{-2mm}
\begin{equation}\label{HSR1}
    R^\mu= \partial_\nu\partial_\lambda h^{\mu\nu\lambda}\, , \quad
    \partial_\mu R_\nu+ \partial_\nu R_\mu\approx 0 \,.
    \vspace{-2mm}
\end{equation}
\vspace{-3mm}
\begin{equation}\label{HSR2}
\widetilde{R}{}^\mu=\partial_\nu\partial_\lambda\widetilde{h}{}^{\mu\nu\lambda}\,,\quad
\partial_\mu \widetilde{R}_\nu + \partial_\nu\widetilde{R}_\mu -
\frac{2}{d}\eta_{\mu\nu}\partial\cdot\widetilde{R}\approx 0\,;
\vspace{-1mm}
\end{equation}
Instead of $\partial_\mu R\approx 0$ for UG, for $s=3$ we arrive at
(conformal) Killing eqs.

The general solution to  eqs. (\ref{HSR1}), (\ref{HSR2}) reads
\vspace{-2mm}
\begin{equation}\label{HSLambda}
    R_\mu=\Lambda_\mu +\Lambda_{\mu\nu}x^\nu\, , \quad
    \widetilde{R}_\mu = \Lambda_\mu +\Lambda_{\mu\nu}x^\nu + \Lambda x^\mu +\Lambda'_\nu(2x^\nu x_\mu-\delta_\mu^\nu x_\lambda x^\lambda)\,,
    \vspace{-2mm}
\end{equation}
where $\Lambda_{\mu\nu}=-\,\Lambda_{\nu\mu}, \Lambda'_\mu,
\Lambda_\mu, \Lambda$ are arbitrary ``integration constants'', being the higher spin analogs of the cosmological constant for the UG.

For $s>3$, the higher Ricci's $R_{\mu_1\dots\mu_{s-2}}$ of the rank $s-2$, or traceless
$\widetilde{R}{}_{\mu_1\dots\mu_{s-2}}$ obey (conformal) Killing
tensor eqs., as  the differential consequences of EoM's. The rank
$s-2$ (conformal) Killing tensor  is decomposed into the product of
$s-2$ (conformal) Killing vectors. Therefore, the number of the
``cosmological constants'' is $10\times(s-2)$ for the
``Maxwell-like'' HS theory, and $15\times(s-2)$ for the UG-like HS
gravity in $d=4$.

Notice that all the theories with unfree gauge symmetry admit alternative description by reducible gauge symmetry with unconstrained gauge parameters \cite{Francia:2013sca}.
Let us consider some examples of such alternative parametrization.

For massless spin 2 in $d=4$,
\vspace{-2mm}
\begin{equation}\label{DeRham}
   \partial_\mu\epsilon^\mu=0 \quad <\approx> \quad  \epsilon^\mu=\partial_\nu\epsilon^{\mu\nu}\,, \quad
   \epsilon^{\mu\nu}=-\,\epsilon^{\nu\mu}\,.
   \vspace{-2mm}
\end{equation}
Equivalence is modulo (Hodge dualised) De Rham cohomology. This form of the  volume preserving diffeomorphism is a  reducible
gauge symmetry. Gauge transformations of gauge parameters read
\vspace{-2mm}
\begin{equation}\label{DR2}
\delta_\omega\epsilon^{\mu\nu}=\varepsilon^{\mu\nu\lambda\rho}\partial_\lambda\omega_{\rho}\,, \quad \delta_\eta \omega^\mu=\partial^\mu \eta\,.
\vspace{-2mm}
\end{equation}
For the ``Maxwell-like'' $s=3$ in $d=4$,
\vspace{-2mm}
\begin{equation}
\partial_\nu \epsilon^{\mu\nu}=0\,, \quad \epsilon^{\mu\nu}=\epsilon^{\nu\mu} \quad <\approx> \quad \epsilon^{\mu\nu}=\partial_\lambda\partial_\rho \epsilon^{\mu\nu\lambda\rho}\,,
\vspace{-2mm}
\end{equation}
where $\epsilon^{\mu\nu\lambda\rho}=\epsilon^{\nu\mu\lambda\rho}$, $\epsilon^{\mu\nu\lambda\rho}=\epsilon^{\mu\nu\rho\lambda}$. Gauge symmetry is reducible,
\vspace{-2mm}
\begin{equation}
    \delta_\omega \epsilon^{\mu\nu\lambda\rho}=\partial_\sigma\omega^{\mu\nu\lambda\rho\sigma}\,, \quad
    \delta_\eta \omega^{\mu\nu\lambda\rho\sigma}=\partial_
    \tau \eta^{\mu\nu\lambda\rho\sigma\tau}\,,
    \vspace{-2mm}
\end{equation}
with gauge parameters of the following symmetry type:
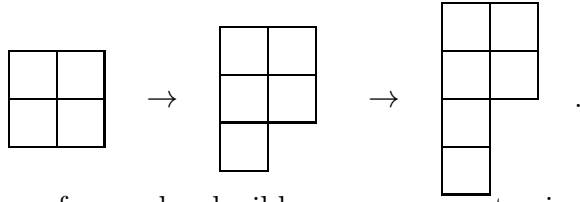
\begin{center}
\ytableausetup{centertableaux}\ytableausetup{centertableaux}
\begin{ytableau}
\phantom{1} & \phantom{1}  \\  \phantom{1} & \phantom{1} \\
\end{ytableau}\,\,\quad $\rightarrow$\,\,\quad
\begin{ytableau}
\phantom{1} & \phantom{1}  \\  \phantom{1}& \phantom{1}   \\ \phantom{1}
\end{ytableau} \,\,\quad$\rightarrow$\,\,\quad
\begin{ytableau}
\phantom{1} & \phantom{1} \\ \phantom{1} & \phantom{1} \\  \phantom{1} \\ \phantom{1}
\end{ytableau}\quad \,.
\end{center}
\vspace{-2mm}
For the connection between unfree and reducible gauge symmetry in Hamiltonian formalism, see \cite{Abakumova:2021mun}.

As one can see from the examples, the dynamics with unfree gauge symmetry do not fit in the usual formalism of gauge systems.  Below, we briefly explain the modifications of the general gauge theory which cover the case of unfree gauge symmetry.

\section{General setup for unfree gauge symmetry, unfree gauge algebra}
Consider Lagrangian field equations
\vspace{-2mm}
\begin{equation}\label{EoM}
    \partial_i S(\phi)\approx  0\,.
    \vspace{-2mm}
    \end{equation}
Proceeding from the observations noticed  in the examples, we
assume the action  $S(\phi)$ to obey \emph{modified  Noether identities} \cite{Kaparulin:2019quz}, \cite{Kaparulin:2019gsx}:
\vspace{-2mm}
\begin{equation}\label{GI}
    \Gamma^i_\alpha\partial_i S +\Gamma^a_\alpha\tau_a\equiv 0\,,
    \vspace{-2mm}
\end{equation}
where $\Gamma$'s are matrices of differential operators, $\tau$ are local quantities. Operator  $\Gamma^a_\alpha$ has \emph{a finite kernel},
\vspace{-2mm}
\begin{equation}\label{KerGamma}
    \Gamma^a_\alpha u_a= 0\quad\Rightarrow
    \quad u\in K\,, \quad \text{dim}\,K=k\in
    \mathbb{N}\,.
    \vspace{-2mm}
\end{equation}
Relations (\ref{GI}), (\ref{KerGamma}) are replace the common definition of gauge symmetry (\ref{GS}) to account for the unfree gauge variation.

Let us explain now, the natural relation of ``the global conserved quantities'' and unfree gauge symmetry.
Once the kernel is finite, elements of $K$ are parameterised by $k$ independent constants $\Lambda_I$, 
\vspace{-2mm}
\begin{equation}\label{LambdaI}
\forall\, u\in K\quad \Rightarrow \quad u=u^I\Lambda_I,\quad I=1\ldots k \, .
\vspace{-2mm}
\end{equation}
The quantities $\tau_a$ are assumed off-shell independent, while on-shell they reduce to elements of $K$, because of (\ref{GI}):
\vspace{-2mm}
\begin{equation}\label{tau-u}
\mathcal{T}_a(\phi,\Lambda)\equiv\tau_a(\phi)- u_a(\Lambda)\approx 0, \quad u_a(\Lambda)\in K
\,.
\vspace{-2mm}
\end{equation}
These relations can be resolved w.r.t. the constants:
\vspace{-2mm}
\begin{equation}\label{JI}
J_I(\phi)\approx \Lambda_I\,,
\vspace{-2mm}
\end{equation}
that means $J_I$ are the \emph{global conserved quantities}. The constants $\Lambda$ are understood as \emph{modular parameters} of the fields. Specific values of $\Lambda$'s are defined by the field asymptotics, or finite number of derivatives at fixed space-time point rather than by Cauchy data.

The local $\Lambda$-dependent  quantities $\mathcal{T}_a(\phi,\Lambda)$ vanish on-shell, while they do not reduce to the linear combinations of EoM's:
\vspace{-2mm}
\begin{equation}
\mathcal{T}_a(\phi,\Lambda)=\tau_a(\phi) - u_a(\Lambda)\approx 0 \, , \quad \mathcal{T}_a\neq \Theta^i_a\partial_i S \,.
\vspace{-2mm}
\end{equation}
These quantities are termed
\emph{completion functions}.

The modified Noether identity (\ref{GI}) means $S(\phi)$ is invariant under
gauge transformations
\vspace{-2mm}
\begin{equation}\label{GT}
\delta_\epsilon \phi^i=\Gamma^i_\alpha\epsilon^\alpha\,,
\vspace{-2mm}
\end{equation}
provided for the gauge parameters $\epsilon^\alpha$ are restricted by equations
\vspace{-2mm}
\begin{equation}\label{GPC}
\Gamma^a_\alpha\epsilon^\alpha=0\,.
\vspace{-2mm}
\end{equation}
With this regard, $\Gamma^a_\alpha$ are termed \emph{gauge parameter constraint operators}.

For gauge symmetry with unrestricted  parameters, any on-shell trivial quantity reduces to linear combination of EoM's, while the gauge parameters are unrestricted. Commutation relations between gauge transformations, and the higher structure relations of gauge algebra, are deduced from Noether identities (\ref{GS}) \cite{Henneaux:1992ig}. In the case of unfree gauge symmetry (\ref{GI}), (\ref{KerGamma}),
any on-shell trivial quantity reduces to linear combination of EoM's \emph{\textbf{and} completion functions} $\mathcal{T}_a(\phi,\Lambda)$. The gauge parameters $\epsilon^\alpha$ are \emph{restricted by the equations} (\ref{GPC}).

Structure relations of unfree gauge symmetry algebra follow from modified Noether identities (\ref{GI}), (\ref{KerGamma}), and they involve, besides gauge generators and EoM's, also completion functions $\tau_a$ and gauge parameter operators $\Gamma^a_\alpha$.

Proceeding from modified Noether identities, with appropriate regularity assumptions for the generators and completion functions \cite{Kaparulin:2019quz}, \cite{Kaparulin:2019gsx}, we arrive at the structure relations involving gauge generators and completion functions:
\begin{equation}\label{Gtau}
    \Gamma^i_\alpha\partial_i\tau_a=R^i_{\alpha a}\partial_iS+R^b_{\alpha a}\tau_b+
    W_{ab}\Gamma^b_\alpha\,;
\end{equation}
\begin{equation}\label{GG}
    \Gamma^i_\alpha\partial_i\Gamma^j_\beta-\Gamma^i_\beta\partial_i\Gamma^j_\alpha=U^\gamma_{\alpha\beta}\Gamma^j_\gamma+E^{aj}_{\alpha\beta}\tau_a+E^{ij}_{\alpha\beta}\partial_i S+R^j_{\alpha a}\Gamma^a_\beta-R^j_{\beta a}\Gamma^a_\alpha\,;
\end{equation}
\begin{equation}\label{GGa}
    \Gamma^i_\alpha\partial_i\Gamma^a_\beta-\Gamma^i_\beta\partial_i\Gamma^a_\alpha=U^\gamma_{\alpha\beta}\Gamma^a_\gamma+R^a_{\alpha b}\Gamma^b_\beta-R^a_{\beta b}\Gamma^b_\alpha
+E^{ab}_{\alpha\beta}\tau_b+E^{ai}_{\alpha\beta}\partial_iS\,,
\end{equation}
where the structure coefficient $W_{ab}$ is on-shell symmetric, and the structure functions $E$ are antisymmetric, $E^{ij}_{\alpha\beta}=-\,E^{ji}_{\alpha\beta}$, $E^{ab}_{\alpha\beta}=-\,E^{ba}_{\alpha\beta}$.

Relation (\ref{Gtau}) means the completion functions are on-shell invariant under unfree gauge variation; (\ref{GG}) demonstrates possible off-shell disclosure of the composition of gauge transformations, including deviation of the parameters from the equations restricting them; and relation (\ref{GGa}) demonstrates that equations imposed on gauge parameters are gauge invariant under unfree gauge variation.

\section{Faddev-Popov (FP) action for unfree gauge symmetry, BV-BRST formalism}

Given the distinctions of the unfree gauge symmetry algebra from the case with unrestricted gauge parameters, the quantisation has to be correspondingly modified. Let us consider  the modification at the level of FP recipe \cite{Kaparulin:2019quz}.

Impose independent gauges $\chi^I(\phi)$, the FP matrix is rectangular,
\vspace{-2mm}
\begin{equation}
\displaystyle \frac{\delta_\varepsilon \chi^I}{\delta\varepsilon^\alpha}=\Gamma^i_\alpha(\phi)\partial_i\chi^I(\phi)\,.
\vspace{-2mm}
\end{equation}
The number of gauges plus the number of equations restricting gauge parameters equals to the number of gauge parameters. The unfree gauge variation has to be transverse to the gauge condition surface.

FP ghosts are introduced being restricted by the equations
\vspace{-2mm}
\begin{equation}
\displaystyle \Gamma^a_\alpha(\phi)C^\alpha=0\,, \quad \text{gh}(C^\alpha)=1\,, \quad \epsilon(C^\alpha)=1\,,
\vspace{-2mm}
\end{equation}
where $\Gamma^a_\alpha(\phi)$ are operators of gauge parameter constraints.

Anti-ghosts are introduced for gauges and equations imposed on ghosts:
\vspace{-2mm}
\begin{equation}
\displaystyle \text{gh}(\bar{C}_I)=\text{gh}(\bar{C}_a)=-1\,, \,\, \epsilon(\bar{C}_I)=\epsilon(\bar{C}_a)=1\,, \,\, \text{gh}(\pi_I)=\epsilon(\pi_I)=0\,.
\vspace{-2mm}
\end{equation}
The FP path integral is adjusted to the case of unfree gauge symmetry:
\vspace{-2mm}
\begin{equation}\label{ZFP}
\displaystyle Z=\int[d\Phi]\exp\Big\{\frac{i}{\hbar}S_{FP}(\phi)\Big\}\,,
\quad \Phi=\{\phi,\pi_i,C^\alpha,\bar{C}_I,\bar{C}_a\}\,,
\vspace{-2mm}
\end{equation}
where the FP action reads
\vspace{-2mm}
\begin{equation}\label{FP}
\displaystyle S_{\text{FP}}=S(\phi)+\pi_I\chi^I(\phi)+\bar{C}_I\Gamma^i_\alpha(\phi)\partial_i\chi^I(\phi)C^\alpha+\bar{C}_a\Gamma^a_\alpha(\phi)C^\alpha\,.
\vspace{-2mm}
\end{equation}
Path integral (\ref{ZFP}) remains unchanged under variation of gauge $\chi$ in the action (\ref{FP}), see in \cite{Kaparulin:2019quz}.
Notice that even for the UG, the FP receipt has been know only for special gauge conditions \cite{Percacci:2017fsy} such that lead to a non-local action, while (\ref{FP}) works well for any local gauge.

The starting point of the BV-BRST formalism extension to the unfree gauge symmetry is the idea that ghosts are constrained
\vspace{-2mm}
\begin{equation}\label{GPCBV}
    \Gamma^a_\alpha C^\alpha=0 \,.
    \vspace{-2mm}
\end{equation}
This equation is considered on equal footing with the original EoM's.
The equation is non-Lagrangian, so it has to be assigned with the anti-field $\xi^a$. For introduction of anti-fields in non-Lagrangian BV-BRST formalism, see \cite{Kazinski:2005eb}.

Once eq. (\ref{GPCBV}) is ghost number one, the anti-field is ghost number zero!
All the fields, including original ones, ghosts, and anti-fields $\xi$ are equipped with anti-canonical conjugate.
The grading is arranged in the Table 1.
\begin{table}[ht]
\caption{}
\begin{center}
\begin{tabular}{| c | c | c | c | c | c | c |}
\hline
\phantom{@}\phantom{@} & \phantom{@}$\phi^i$\phantom{@} & \phantom{1} $\xi^a$  \phantom{1}& \phantom{'} $C^\alpha$ \phantom{'} & \phantom{@}$\phi^*_i$\phantom{@} &\phantom{@}$\xi^*_a$\phantom{@} & \phantom{@}$C^*_\alpha$\phantom{@}\\ \hline
$\varepsilon$ & $0$ & $0$  & $1$ & $1$ & $1$ & $0$\\ \hline
gh & $0$ & $0$  & $1$ & $-\,1$ & $-\,1$ & $-\,2$\\ \hline
deg & $0$ & $1$  & $0$ & $1$ & $1$ & $2$\\ \hline
\end{tabular}
\end{center}
\end{table}

Given the anti-canonical pairs, the anti-bracket reads
\vspace{-2mm}
\begin{equation}
    (A,B)=\frac{\partial^R A}{\partial \varphi^I}\frac{\partial^L B}{\partial \varphi^*_I}-\frac{\partial^R A}{\partial \varphi^*_I}\frac{\partial^L B}{\partial \varphi^I}\,,
    \vspace{-2mm}
\end{equation}
where $\varphi^I=(\phi^i,\xi^a,C^\alpha)$, $\varphi^*_I=(\phi^*_i,\xi^*_a,C^*_\alpha)
$, and
\vspace{-2mm}
\begin{equation}
\text{gh}((A,B))=\text{gh}(A)+\text{gh}(B)+1\,, \quad
\varepsilon((A,B))=\varepsilon(A)+\varepsilon(B)+1\,.
\vspace{-2mm}
\end{equation}
The BV action is defined by the master equation
\vspace{-2mm}
\begin{equation}\label{BVME}
(S,S)=0 \,.
\vspace{-2mm}
\end{equation}
The solution is sought for as the expansion w.r.t. resolution degree
\vspace{-2mm}
\begin{equation}
S=\sum\limits_{k=0}S_k\,, \quad \text{gh}(S_k)=\varepsilon(S_k)=0\,, \quad \text{deg}\,S_k=k\,.
\vspace{-2mm}
\end{equation}
The boundary condition is defined by the first two orders
\vspace{-2mm}
\begin{equation}
    \displaystyle S_0=S(\phi), \quad S_1=\tau_a\xi^a+(\phi^*_i\Gamma^i_\alpha+\xi^*_a\Gamma^a_\alpha)C^\alpha\,,
    \vspace{-2mm}
\end{equation}
where $S$ is the original action, while $S_1$ includes the basic constituents of unfree gauge symmetry: completion functions $\tau_a$, gauge generators $\Gamma^i_\alpha$, and operators of gauge parameter constraints $\Gamma^a_\alpha$.
The second order reads
\vspace{-2mm}
\begin{equation}
\begin{array}{c}
\displaystyle S_2=\frac{1}{2}(C^*_\gamma U^{\gamma}_{\alpha\beta}+\phi^*_j\phi^*_iE^{ij}_{\alpha\beta}+2\xi^*_a\phi^*_iE^{ia}_{\alpha\beta}+\xi^*_b\xi^*_aE^{ab}_{\alpha\beta})C^\alpha C^\beta\\
\displaystyle-\,\xi^b(\phi^*_iR^i_{b\alpha}+\xi^*_aR^a_{b\alpha})C^\alpha-\frac{1}{2}\xi^b\xi^aW_{ab}\,.
\end{array}
\vspace{-2mm}
\end{equation}
Master equation (\ref{BVME}) identifies all the coefficients in $S_2$ with structure functions in structure relations (\ref{Gtau})--(\ref{GGa}) of unfree gauge symmetry algebra.

BRST differential $s$ is anti-Hamiltonian vector field for the master action:
\vspace{-2mm}
\begin{equation}
sA=(A,S)\,, \quad s^2=0\,, \quad \text{gh}(s)=1\,, \quad \varepsilon(s)=1\,.
\vspace{-2mm}
\end{equation}
It can be decomposed w.r.t. resolution degree
\vspace{-2mm}
\begin{equation}
s=\delta+\gamma+\overset{(1)}{s}+\ldots\,, \,\, \text{deg}\,\delta=-\,1\,, \,\, \text{deg}\,\gamma=0\,, \,\, \text{deg}\,\overset{(1)}{s}=1\,.
\vspace{-2mm}
\end{equation}
Because of master equation, the first orders are connected by the relations
\vspace{-2mm}
\begin{equation}\label{delta-gamma}
s^2=0\,\,\Rightarrow\,\, \delta^2=0\,, \,\, \delta\gamma+\gamma\delta=0\,,
\,\,\gamma^2+(\delta\overset{(1)}{s}+\overset{(1)}s\delta)=0\,,
\vspace{-2mm}
\end{equation}
where Kozul-Tate differential $\delta$ is defined as
\vspace{-2mm}
\begin{equation}
\delta A=-\,\frac{\partial^R A}{\partial\phi^*_i}\partial_iS-\frac{\partial^R A}{\partial C^*_\alpha}(\phi^*_i\Gamma^i_\alpha+\xi^*_a\Gamma^a_\alpha)+\frac{\partial^R A}{\partial\xi^a}\Gamma^a_\alpha C^\alpha\,.
\vspace{-2mm}
\end{equation}
By virtue of Noether identity for unfree gauge symmetry, $\delta$ squares to zero,
\vspace{-2mm}
\begin{equation}
\delta^2 A=-\,\frac{\partial^R A}{\partial C^*_a}(\Gamma^i_\alpha\partial_iS+\Gamma^a_\alpha\tau_a)\equiv 0\,.
\vspace{-2mm}
\end{equation}
One can verify that $\delta$ is acyclic in strictly positive resolution degrees, that insures existence of solution for $s$ in the $\text{deg}>0$, Q.E.D. For more details, see \cite{Kaparulin:2019gsx}.

Given the extension of the BV formalism to the case of unfree symmetry, one can seek for consistent deformations of the models of this class and systematically quantize them.

\section{Unfree gauge symmetry in Hamiltonian formalism}

Hamiltonian action for the theory with primary constraints $T_\alpha$ reads
\vspace{-2mm}
\begin{equation}\label{S-H}
S_H=\int dt \big(p_i\dot{q}^i-H_T\big)\,, \,\, H_T=H+\lambda^\alpha T_\alpha\,,
\vspace{-2mm}
\end{equation}
where the role of fields is played by canonical variables $q^i$, $p_i$, and Lagrange multipliers $\lambda^\alpha$.
Assume that there are no second-class constraints. Conservation of $T_\alpha$ leads to secondary constraints $\tau_a$,
\vspace{-2mm}
\begin{equation}
\dot{T}_\alpha\equiv\{T_\alpha\,, H_T\}=W_\alpha^\beta T_\beta(q,p)+\Gamma_\alpha^a\tau_a(q,p)\approx0\,,
\vspace{-2mm}
\end{equation}
where $W, \Gamma$ are local differential operators, $\Gamma$ has \emph{finite kernel}. Secondary constraints $\tau$ are considered as \emph{completion functions}, and gauge symmetry should be unfree. Once the kernel of $\Gamma$ is finite, completion functions can be redefined by adding modular parameters $\Lambda$ to make $\tau$ vanishing on-shell,
\vspace{-2mm}
\begin{equation}
\Gamma_\alpha^a\tau_a=0 \,\,\Leftrightarrow \,\,\tau_a=\Lambda_a\,, \, \,\Lambda_a \in \text{Ker}\,\Gamma_\alpha^a\,:\, \tau_a \mapsto \tau_a-\Lambda_a\,.
\vspace{-2mm}
\end{equation}
Assume no tertiary constraints appear,
\begin{equation}
\dot{\tau}_a\equiv\{\tau_a\,,H_T\}=W_a^\alpha T_\alpha(q,p)+W_a^b \tau_b(q,p)
\approx 0\,.
\end{equation}
For more general case, see \cite{Abakumova:2020ajc}.

 Termination of the Dirac-Bergmann algorithm means the modified gauge identities as the EoM's turn out dependent with their differential consequences \emph{and completion functions}:
 \vspace{-2mm}
\begin{equation}
\begin{array}{c}
\displaystyle \{T_\alpha\,,q^i\}\frac{\delta S_H}{\delta q^i}+\{T_\alpha\,,p_i\}\frac{\delta S_H}{\delta p_i}
+\big(\delta_\alpha^\beta\frac{d}{dt}-W_\alpha^\beta\big)\frac{\delta S_H}{\delta\lambda^\beta}
+\Gamma_\alpha^a\tau_a\equiv0\,;
\\[2mm]
\displaystyle \{\tau_a\,,q^i\}\frac{\delta S_H}{\delta q^i}+\{\tau_a\,,p_i\}\frac{\delta S_H}{\delta p_i}
-W_a^\alpha\frac{\delta S_H}{\delta\lambda^\alpha}
+\big(-\delta_a^b\frac{d}{dt}+W_a^b\big)\tau_b\equiv0\,.
\end{array}
\vspace{-2mm}
\end{equation}
Corresponding unfree gauge symmetry transformations read
\vspace{-2mm}
\begin{equation}\label{GT-H}
\begin{array}{c}
\delta_\varepsilon O(q,p)=\{O\,,T_\alpha\}\varepsilon^\alpha+\{O\,,\tau_a\}\varepsilon^a\,,\\[1mm]
\delta_\varepsilon \lambda^\alpha=\dot{\varepsilon}^\alpha+W^\alpha_\beta\varepsilon^\beta+W_a^\alpha\varepsilon^a\,.
\vspace{-2mm}
\end{array}
\end{equation}
Constraints on gauge parameters take the form
\vspace{-2mm}
\begin{equation}\label{GC-H}
\big(\delta_b^a\frac{d}{dt}+W_b^a\big)\varepsilon^b+\Gamma_\alpha^a\varepsilon^\alpha=0\,.
\vspace{-2mm}
\end{equation}
Direct computation confirms that action (\ref{S-H}) is invariant under transformations (\ref{GT-H}), (\ref{GC-H}),
\vspace{-2mm}
\begin{equation}
\delta_\varepsilon S_H\equiv\int dt \big[\big((\delta_b^a\frac{d}{dt}+W_b^a)\varepsilon^b+\Gamma_\alpha^a\varepsilon^\alpha\big)\tau_a-\frac{1}{2}\frac{d}{dt}\big(T_\alpha\varepsilon^\alpha+\tau_a\varepsilon^a\big)\big]=0\,.
\vspace{-2mm}
\end{equation}

For the linearised unimodular gravity (LUG), Hamiltonian action (\ref{S-H}) reads
\vspace{-2mm}
\begin{equation}
\begin{array}{l}
\displaystyle S_H[h,\Pi, \lambda]=\int d^4x\big(\Pi^{ij}\dot{h}_{ij}-H-\lambda^i T_i\big)\,, \quad
T_i=-\,2\partial^j\Pi_{ij}\,,
\\[2mm]
\displaystyle H=\Pi^{ij}\Pi_{ij}-\frac{1}{2}\Pi^2+\frac{1}{4}\big(2\partial^i h_{ij} \partial_k h^{kj}-\partial_i h\partial^i h-\partial_i h_{jk} \partial^i h^{jk}\big)\,,
\end{array}
\vspace{-2mm}
\end{equation}
where $i,j,k=1,2,3$, $h=\eta^{ij}h_{ij}$, $\Pi=\eta_{ij}\Pi^{ij}$, $\lambda^i=h^{0i}$.

Conservation of primary constraints $T_i$ leads to the secondary ones,
\vspace{-2mm}
\begin{equation}
\displaystyle \dot{T}_i=\{T_i,H\}=-\,\partial_i\tau_0=0\,, \quad \tau_0\equiv\partial^i\partial^jh_{ij}-\partial_i\partial^ih-\Lambda=0\,.
\vspace{-2mm}
\end{equation}
The secondary constraints are conserved by virtue of the primary ones:
\vspace{-2mm}
\begin{equation}
\displaystyle \dot{\tau}_0=\{\tau_0,H\}=-\,\partial^i T_i\,.
\vspace{-2mm}
\end{equation}
Unfree gauge symmetry transformations read
\vspace{-2mm}
\begin{equation}
\displaystyle \delta_\varepsilon h_{ij}=\partial_i\varepsilon_j+\partial_j\varepsilon_i\,, \,\, \delta_\varepsilon \Pi^{ij}=-\,\partial^i\partial^j\varepsilon^0+\eta^{ij}\partial_k\partial^k\varepsilon^0\,,
\,\, \delta_\varepsilon \lambda^i=\dot{\varepsilon}{}^i+\partial^i\varepsilon^0\,.
\vspace{-2mm}
\end{equation}
Gauge variation of the action reads
\vspace{-2mm}
\begin{equation}
\displaystyle \delta_\varepsilon S_H\equiv\int d^4x\big((\dot{\varepsilon}{}^0+\partial_i\varepsilon^i)\tau_0-\partial_0(T_i\varepsilon^i+\tau_0\varepsilon^0)\big)\,.
\vspace{-2mm}
\end{equation}
So, gauge parameters have to obey equation
\vspace{-2mm}
\begin{equation}
\displaystyle \dot{\varepsilon}{}^0+\partial_i\varepsilon^i=0\,.
\vspace{-2mm}
\end{equation}
For more detailed description, see \cite{Abakumova:2019uoo}. For analogue in the non-linear UG, see \cite{Karataeva:2022mll}.

\section{Hamiltonian BFV-BRST formalism}

To avoid technical complexities, we restrict consideration by simplified involution relations
\vspace{-2mm}
\begin{equation}
\begin{array}{c}
\{T_\alpha, H\}=V_\alpha^a\tau_a\,, \,\, \{\tau_a, H\}=V_a^\alpha T_\alpha\,,
\\[1mm]
\{T_\alpha,T_\beta\}=\{T_\alpha,\tau_a\}=\{\tau_a,\tau_b\}=0\,,
\end{array}
\vspace{-2mm}
\end{equation}
with structure coefficients $V^a_\alpha$, $V^\alpha_a$ being constants.

Complete BRST charge reads
\vspace{-2mm}
\begin{equation}
\displaystyle Q=T_\alpha C^\alpha+\tau_a C^a+\pi_ \alpha P^\alpha
\vspace{-2mm}
\end{equation}
Given the gauge conditions,
\vspace{-2mm}
\begin{equation}
\displaystyle \dot{\lambda}^\alpha-\chi^\alpha=0\,,
\vspace{-2mm}
\end{equation}
the gauge fermion is introduced,
\vspace{-2mm}
\begin{equation}
\displaystyle \Psi=\bar{C}_\alpha \chi^\alpha+\lambda^\alpha\overline{P}_\alpha\,,
\vspace{-2mm}
\end{equation}
and gauge-fixed Hamiltonian is defined by the usual rule,
\vspace{-2mm}
\begin{equation}
\begin{array}{c}
H_{\Psi}=\mathcal{H}+\{Q,\Psi\}=H-\overline{P}_\alpha V^\alpha_aC^a-\overline{P}_aV^a_\alpha C^\alpha+T_\alpha\lambda^\alpha+\pi_\alpha\chi^\alpha
\\[1mm]
+\,\overline{P}_\alpha P^\alpha+\overline{C}_\alpha\{\chi^\alpha,T_\beta\}C^\beta+\overline{C}_\alpha\{\chi^\alpha,\tau_a\}C^a\,.
\end{array}
\vspace{-2mm}
\end{equation}
The grading is arranged in the Table 2.
For more detailed description, see \cite{Abakumova:2020ajc}, \cite{Abakumova:2021mun}, \cite{Abakumova:2021rlq}.

\begin{table}[ht]
\caption{}
\begin{center}
\begin{tabular}{| c | c | c | c | c | c | c | c | c | c |}
\hline
$\overset{\phantom{}}{\phantom{@}}$\phantom{@} & \phantom{''}$C^\alpha$\phantom{''} & \phantom{'} $\overline{P}_\alpha$  \phantom{'}& \phantom{''} $C^a$ \phantom{' '} & \phantom{'}$\overline{P}_a$\phantom{'} &\phantom{''}$\lambda^\alpha$\phantom{''} & \phantom{''}$\pi_\alpha$\phantom{''} & \phantom{''}$P^\alpha$\phantom{''} & \phantom{'}$\overline{C}_\alpha$\phantom{'} \\ \hline
$\varepsilon$ & $1$ & $1$  & $1$ & $1$ & $0$ & $0$ & $1$ & $1$ \\ \hline
gh & $1$ & $-\,1$  & $1$ & $-\,1$ & $0$ & $0$  & $1$ & $-\,1$ \\ \hline
\end{tabular}
\end{center}
\end{table}

For the UG, the complete BRST charge reads \cite{Karataeva:2022mll}
\begin{equation}\label{Q-UG}
\begin{array}{c}
\displaystyle Q=\int d^3 x\big(T_i C^i+\tau C-\overline{P}_i C^j\partial_j C^i-\overline{P}\partial_i(C^iC)
\\[1mm]
\displaystyle +\,\overline{P}_i(-\,\overset{*}g){}^{-\,1}\overset{*}{g}{}^{ij}C\partial_jC+\pi_i P^i\big)\,, \quad
T_i=-\,2\overset{*}{g}_{ij}\big(\partial_k\Pi^{kj}+\overset{*}{\Gamma}{}^j_{kl}\Pi^{kl}\big)\,,
\\[1mm]
\displaystyle \tau=H-\Lambda=-\,\frac{1}{\overset{*}{g}}\mathcal{G}_{ijkl}\Pi^{ij}\Pi^{kl}+\overset{*}{R}-\Lambda\,, \quad \Lambda=const\,.
\end{array}
\end{equation}
Introducing the gauge fermion
\begin{equation}
\Psi=\int d^3x \big(\overline{C}_i\chi^i+\overline{P_i}N^i\big)\,, \quad \chi^i=(-\,\overset{*}{g}){}^{-\,1}\partial_j\overset{*}{g}{}^{ji}+N^j\partial_j N^i\,,
\end{equation}
we arrive to the complete gauge-fixed BRST-invariant Hamiltonian
\begin{equation}
\begin{array}{c}
\displaystyle H_{\Psi}=\mathcal{H}+\{Q,\Psi\}=\int d^3x \big\{H+T_iN^i+\pi_i\chi^i-\overline{P}\partial_iC^i\\[1mm]
\displaystyle-\,\overline{P}_i(-\,\overset{*}{g}){}^{-\,1}\overset{*}{g}{}^{ij}\partial_jC+\overline{P}\partial_i(CN^i)+\overline{P}_i(\partial_jC^iN^j-C^j\partial_jN^i)\\[1mm]
\displaystyle -\,\overline{C}_i(-\,\overset{*}g){}^{-\,1}\big(2\partial_j\overset{*}{g}{}^{ij}\overset{*}{\nabla}_kC^k+\partial_j(\overset{*}{\nabla}{}^jC^i+\overset{*}{\nabla}{}^iC^j)\big)\\[1mm]
\displaystyle +\,\overline{C}_i(-\,\overset{*}{g}){}^{-\,1}\big(\partial_j\overset{*}{g}{}^{ij}(-\,\overset{*}{g}){}^{-\,1}\Pi C-2\partial_j((-\,\overset{*}{g}){}^{-\,1}\Pi^{ij}C)\\[1mm]
\displaystyle +\,\partial_j(\overset{*}{g}{}^{ij}(-\,\overset{*}{g}){}^{-\,1}\Pi C)\big)
+\overline{C}_i(\partial_jN^iP^j+N^j\partial_jP^i)+\overline{P}_i P^i\big\}\,.
\end{array}
\end{equation}
As one can see, the Hamiltonian does not involve $\Lambda$, nor does it contain four-ghost vertices, unlike the GR analog. The BRST charge (\ref{Q-UG}) involves cosmological constant explicitly. It is included in $Q$ in a manner similar to inclusion of the intercept in the bosonic string theory. As the theory is non-renormalisable, discussion of the consequences of this specifics is somewhat speculative. If the way was known to make serious quantum calculations, this could mean that the spectrum of admissible physical states is sensitive to $\Lambda$, and $\Lambda$ can be even quantized, getting spectrum. On the other hand, no quantum transitions are possible between the states with different $\Lambda$, as it is involved in $Q$ as a modular parameter. If one consider an alternative reducible parameterisation of gauge symmetry as it is done in \cite{Karataeva:2022mll}, this would lead to inequivalent BFV-BRST formalism, where $\Lambda$ is not explicitly involved in BRST charge. From this perspective, cosmological constant is the BRST co-cycle, and quantum transitions can be possible between the states with different $\Lambda$.

\label{sec:concl}
\section*{Conclusion}
Besides action and gauge generators, the unfree gauge symmetry algebra has two more principal constituents: operators of gauge parameter constraints and completion functions.
Noether identities are modified involving these constituents (\ref{GI}), (\ref{KerGamma}). This results in modification of structure relations of gauge algebra.
Modified Noether identities result in the ``global conserved quantities'' in any model with unfree gauge symmetry. The modification is found for the FP ansatz that accounts for the constraints imposed on the gauge parameters. This has consequences in the models, including UG. The BV-BRST field-anti-field formalism is worked out that accounts for the unfree gauge symmetry. The unfree gauge symmetry transformations are described in terms of general constrained Hamiltonian formalism. The volume preserving diffeomorphisms are constructed in Hamiltonian form of UG. Hamiltonian BFV-BRST formalism is worked out for the systems with unfree gauge symmetry. Being applied to the UG, it results in previously unknown ghost vertices in the complete gauge fixed BRST invariant Hamiltonian.

\vspace{2 mm}

\subsection*{Acknowledgments.}
In the studies of unfree gauge symmetry, the authors benefited from collaboration with D. Kaparulin and I. Karataeva. Various aspects of these studies have been discussed with A. Barvinsky, D. Francia, A. Kamenshchik and A. Sharapov. The authors are grateful to all the mentioned colleagues.

The part of the work concerning BV and BFV formalism is supported by Foundation for
Advancement of Theoretical Physics and Mathematics  ``BASIS''. Study of the BRST quantization of the UG is supported by a government task of the Ministry of Science and Higher Education of the Russian Federation, Project No. FSWM-2020-0033.

\end{document}